\UseRawInputEncoding
\pdfoutput=1

\documentclass[%
reprint,
superscriptaddress,
amsmath,amssymb,
aps,
pra,
]{revtex4-2}

\usepackage{graphicx}
\usepackage{dcolumn}
\usepackage{bm}
\begin{document}

\preprint{APS/123-QED}

\title{Quantum-Classical Computational Molecular Design of\\Deuterated High-Efficiency OLED Emitters}

\author{Qi Gao}
 \email{caoch@user.keio.ac.jp}
\affiliation{Mitsubishi Chemical Corporation, Science \& Innovation Center, 1000, Kamoshida-cho, Aoba-ku, Yokohama 227-8502, Japan}
\affiliation{Quantum Computing Center, Keio University, Hiyoshi 3-14-1, Kohoku, Yokohama 223-8522, Japan}
\author{Gavin O. Jones}
 \email{gojones@us.ibm.com}
\affiliation{IBM Quantum, IBM Research – Almaden, 650 Harry Road, San Jose, CA 95120, USA}
\author{Takao Kobayashi}
\affiliation{Mitsubishi Chemical Corporation, Science \& Innovation Center, 1000, Kamoshida-cho, Aoba-ku, Yokohama 227-8502, Japan}
\affiliation{Quantum Computing Center, Keio University, Hiyoshi 3-14-1, Kohoku, Yokohama 223-8522, Japan}
\author{Michihiko Sugawara}
\affiliation{Quantum Computing Center, Keio University, Hiyoshi 3-14-1, Kohoku, Yokohama 223-8522, Japan}
\author{Hiroki Yamashita}
\affiliation{Mitsubishi Chemical Corporation, Science \& Innovation Center, 1000, Kamoshida-cho, Aoba-ku, Yokohama 227-8502, Japan}
\author{Hideaki Kawaguchi}
\affiliation{Quantum Computing Center, Keio University, Hiyoshi 3-14-1, Kohoku, Yokohama 223-8522, Japan}
\author{Shu Tanaka}
\affiliation{Quantum Computing Center, Keio University, Hiyoshi 3-14-1, Kohoku, Yokohama 223-8522, Japan}
\author{Naoki Yamamoto}
 \email{yamamoto@appi.keio.ac.jp}
\affiliation{Quantum Computing Center, Keio University, Hiyoshi 3-14-1, Kohoku, Yokohama 223-8522, Japan}

\date{\today}

\begin{abstract}
This study describes a hybrid quantum-classical computational approach for designing synthesizable deuterated $Alq_3$ emitters possessing desirable emission quantum efficiencies (QEs). This design process has been performed on the tris(8-hydroxyquinolinato) ligands typically bound to aluminum in $Alq_3$. It involves a multi-pronged approach which first utilizes classical quantum chemistry to predict the emission QEs of the $Alq_3$ ligands. These initial results were then used as a machine learning dataset for a factorization machine-based model which was applied to construct an Ising Hamiltonian to predict emission quantum efficiencies on a classical computer. We show that such a factorization machine-based approach can yield accurate property predictions for all 64 deuterated $Alq_3$ emitters with 13 training values. Moreover, another Ising Hamiltonian could be constructed by including synthetic constraints which could be used to perform optimizations on a quantum simulator and device using the variational quantum eigensolver (VQE) and quantum approximate optimization algorithm (QAOA) to discover a molecule possessing the optimal QE and synthetic cost. We observe that both VQE and QAOA calculations can predict the optimal molecule with greater than 0.95 probability on quantum simulators. These probabilities decrease to 0.83 and 0.075 for simulations with VQE and QAOA, respectively, on a quantum device, but these can be improved to 0.90 and 0.084 by mitigating readout error. Application of a binary search routine on quantum devices improves these results to a probability of 0.97 for simulations involving VQE and QAOA.
\end{abstract}

\maketitle



\section{\label{sec:level1}Introduction}

Organic luminescent materials have attracted significant and increasing interest from academia and industry over the past decade, particularly for the fabrication of lightweight and flexible optoelectronic devices which can be used in a variety of applications including OLED displays \cite{zhen-gang_2005, bera_2005}, luminescent solar concentrators \cite{earp_2011, van_sark_2013} and photocatalysts \cite{kang_2021, yu_2020}. While possessing many potential advantages, several serious challenges must be overcome before organic luminescent materials can be commercialized.

One challenge involves the need to improve the emission quantum efficiency, $\Phi$, defined as:
\begin{equation}
\Phi = \frac{\kappa_r} {\kappa_r + \kappa_{nr}}
\label{eq:one}
\end{equation}
where $\kappa_r$ and $\kappa_{nr}$ are the rate constants for all radiative and non-radiative relaxation processes \cite{lakowicz_2006}. Since such materials intrinsically suffer from a variety of non-radiative decay modes, one method to achieve a high emission quantum efficiency is to lower $\kappa_{nr}$. Based on the Fermi golden rule of time-dependent perturbation theory, the $\kappa_{nr}$ between two different electronic states with equal spin multiplicity in the weak nonadiabatic coupling case is proportional to the product of an electronic term of the nonadiabatic coupling, and a vibrational term comprising the Franck-Condon (FC) factor defined by the square of the overlap between total vibrational wavefunctions of different electronic states.

Vibrational normal modes with high vibrational frequencies are required to fulfill the energy conservation law. For example, C-H stretching modes ($\sim$3000 $cm^{-1}$) dominate FC factors since they are widely spaced in energy and possess much smaller vibrational quantum numbers than other vibrational normal modes comprising lower electronic states. Therefore, introduction of heavier deuterium isotopes to replace hydrogen atoms in C-H bonds could effectively lower FC factors ($\kappa_{nr}$) since the vibrational frequencies of C-D stretching modes are on the order of 2200 $cm^{-1}$. Many experimental studies show that substitution of hydrogen by its deuterium isotope improve the luminescent performance of organic materials \cite{sugiyama_2013, sugiyama_2017, yuan_2021, bischof_2010} and could potentially be used for optoelectronic applications such as OLED devices \cite{sugiyama_2013} and image probes \cite{sugiyama_2017}.

To accelerate the design of deuterated organic emitters with high emission quantum efficiencies, it would be quite useful to predict the positional dependence of deuteron substitutions on the FC factor using \textit{ab initio} methods, leading to efficiently deuterated molecules possessing high emission QEs. However, such simulations of FC factors could be computationally demanding since an emitter possessing n hydrogen atoms would require evaluations of $2^n$ possible deuterated molecules. Moreover, as a practical consideration, increasing the amount of deuterium in molecules significantly increases the cost of chemical synthesis \cite{sugiyama_2013, sugiyama_2017}, thus deuterated molecules produced by this design should not only exhibit high quantum efficiency but must also be synthesizable. Previous investigations have shown that machine learning is orders of magnitude faster than \textit{ab initio} calculation for predicting desired properties and synthesizability can also be integrated into a training model \cite{bjerrum_2017, sumita_2018}, therefore, here we propose the use of a machine learning model to accelerate the discovery of optimally deuterated OLED emitters \cite{butler_2018}.

Such a machine learning model, trained to predict molecular properties based on molecular structure, can in principle generate data that can be enhanced by hybrid quantum-classical variational quantum optimization algorithms (VQAs) such as the Quantum Approximate Optimization Algorithm (QAOA) \cite{farhi_2014} and the Variational Quantum Eigensolver (VQE) \cite{peruzzo_2014}. These algorithms are suitable for execution on current noisy quantum hardware since they execute a short, parameterized quantum circuit on the devices while parameters are optimized using a classical outer loop to enable the discovery of high-quality solutions by the quantum circuit.

In this study, we develop a combined quantum chemistry, machine learning and quantum optimization approach to discover an optimal, synthesizable deuterated molecule of $Alq_3$ (a well-known OLED emitter \cite{colle_2002, brinkmann_2000}) possessing high quantum efficiency. Our approach involves a sequence initiated by the evaluation of the Franck-Condon factors of various deuterated $Alq_3$ derivatives with classical computational chemistry methods. This step is followed by the selection of a suitable prediction model via machine learning performed on classical architecture. The sequence terminates with the use of an Ising Hamiltonian derived from this machine learning approach in both constrained and unconstrained quantum optimization procedures involving the use of the VQE and QAOA algorithms to predict the optimal deuterated molecule.

\section{\label{sec:level1}Methods}

\subsection{\label{sec:level2}Computational Methodology}

\begin{figure}
\includegraphics{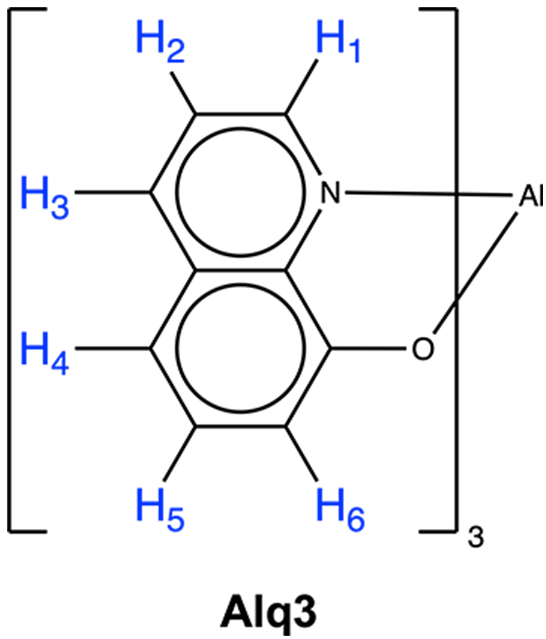}
\caption{\label{fig:fig1}Structure of $Alq_3$ and the numbering of the 6 hydrogen atoms in the 8-hydroxyquinoline ligand.}
\end{figure}

$Alq_3$ (tris(8-hydroxyquinolinato)aluminium) is a stable metal chelate complex wherein aluminum is bonded to three 8-hydroxyquinoline ligands, as shown in Figure 1. For every ligand, there are six hydrogen atoms which can be replaced by deuterium isotopes \cite{sugiyama_2013}. By assuming that every ligand has the same deuterated structure and setting every hydrogen atom and its deuterium as bit values of 1 and 0 respectively, the combinatorial optimization required for discovering a desired deuterated $Alq_3$ molecule with high emission quantum efficiency can be viewed as a problem involving searching for an optimal bitstring represented by 6 qubits on a quantum computer.

\begin{figure}
\includegraphics{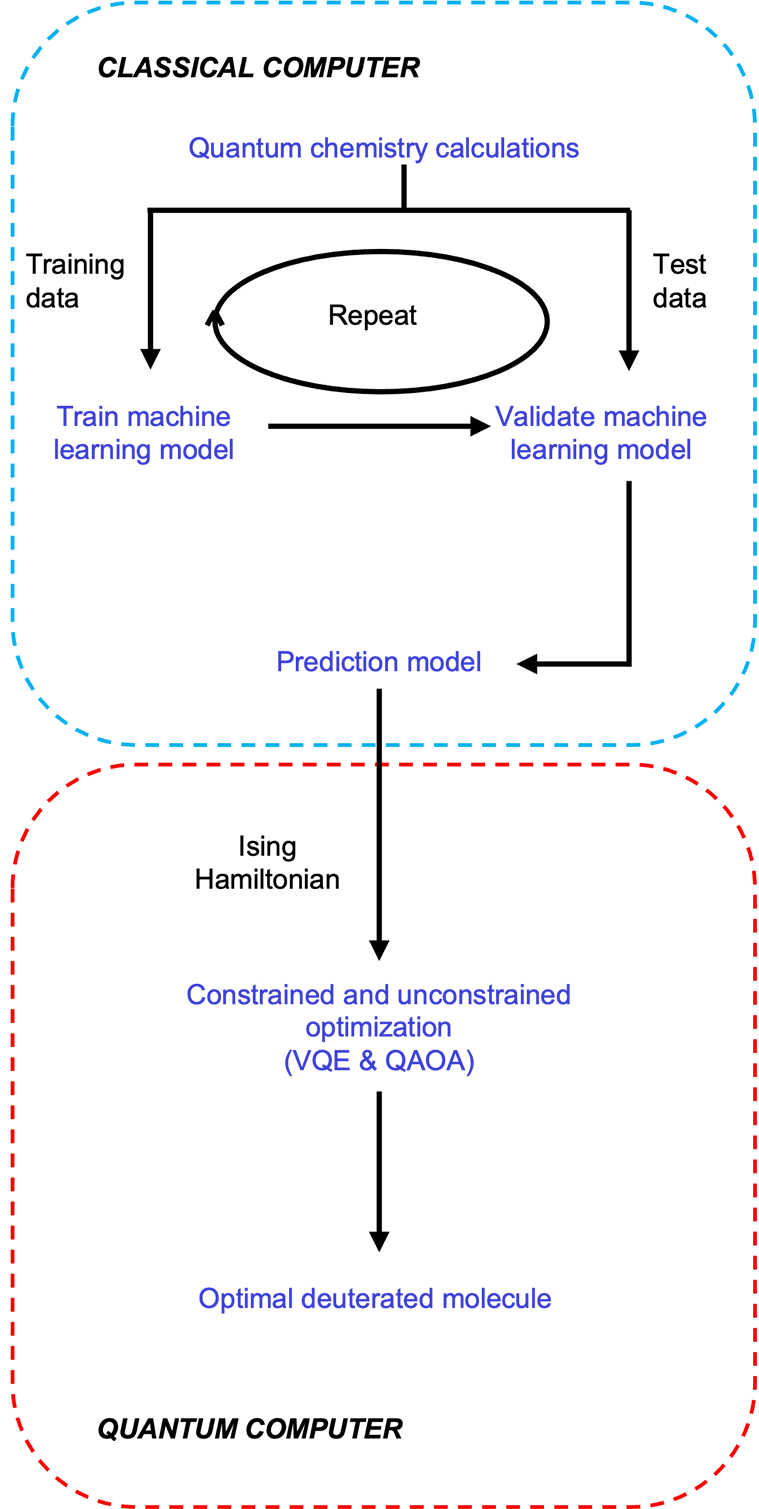}
\caption{\label{fig:fig2}Flowchart for a hybrid quantum-classical quantum chemistry, machine learning and quantum optimization approach for identifying optimally deuterated $Alq_3$ derivatives with high quantum efficiencies.}
\end{figure}

Figure 2 shows a flowchart illustrating the hybrid quantum-classical procedure combining quantum chemistry, machine learning and quantum optimization to identify deuterated $Alq_3$ molecules possessing desired quantum efficiency. This procedure begins with quantum chemistry calculations with a classical computer to obtain the quantum efficiencies of a set of deuterated $Alq_3$ molecules. Results obtained from these calculations are used to prepare a training dataset used to train a model to predict the quantum efficiency, and a test dataset used to validate the performance of the trained model by comparing predicted values with those obtained using the test dataset. This procedure is repeated with additional quantum chemistry results added to the training dataset until the predicted accuracy of the trained model achieves a desired threshold. The improved model is used to construct an Ising Hamiltonian which is used to perform unconstrained and constrained optimization using a quantum computer to find the optimally deuterated $Alq_3$ derivatives. The following sections describe the details of the different aspects of this procedure involving quantum chemistry calculations, machine learning, and quantum optimizations on quantum devices and simulators.

\subsection{\label{sec:level2}Quantum Chemistry and Machine Learning Methodology}

In order to calculate the FC factors for nonradiative decay from $S_1$ to $S_0$ of the variously substituted deuterated $Alq_3$ molecules, the calculations of harmonic vibrational frequencies and normal modes of the structures with geometries optimized on the $S_0$ and $S_1$ electronic states are performed with the CAM-B3LYP \cite{yanai_2004} and (TD)CAM-B3LYP methods \cite{adamo_2013} and the 6-31G(d,p) basis sets using Gaussian09 \cite{frisch_2009}. The FC factors for the $S_1 \rightarrow S_0$ nonradiative decay of non-deuterated and deuterated $Alq_3$ derivatives were calculated from the $S_0$ and $S_1$ vibrational frequencies and normal modes using MOMAP version 0.3.001 \cite{shuai_2020}. These calculations consist of the square of the overlap between the vibrational wavefunctions of the $S_1$ and $S_0$ states with the adiabatic excitation energy of the $S_1$ excited state above the $S_0$ vibrational ground state at $300 K$.

Using the numbering scheme for the hydrogen atoms in the ligand as shown in Figure 1, every deuterated molecular structure is transformed into a binary feature vector $\vec{x}^{(s)}$ represented by 6 qubits. A dataset is prepared with all the $\vec{x}^{(s)}$ and its corresponding target values of $\vec{y}^{(s)}$ as shown in Table 1. The non-deuterated $Alq_3$ molecule (i.e., $\vec{x}^{(s)} = [111111]$) was chosen and a singly deuterated structure was randomly selected as the training data. In addition to this selection, deuterated structures possessing the normal reverse feature vectors (for example, $\vec{x}^{(s)} = [011111]$ and $\vec{x}^{(s)} = [100000]$) of the selected $\vec{x}^{(s)}$ were also used in the training data. The selected training data are used to train a prediction model described by the quadratic unconstrained binary optimization (QUBO) formulation \cite{kochenberger_2014}:

\begin{table}
\caption{\label{tab:table1}%
Dataset obtained from quantum chemistry calculations of deuterated $Alq_3$ molecules for training and validating the machine learning model. The bit values of 0 and 1 in the feature vector $\vec{x}^(s)$ correspond to the deuterium and hydrogen atom in the $Alq_3$ molecule.
}
\begin{tabular}{|p{0.75cm}|p{0.5cm}|p{0.5cm}|p{0.5cm}|p{0.5cm}|p{0.5cm}|p{0.5cm}||p{0.75cm}|p{1.5cm}|}
\hline
\multicolumn{7}{|c||}{\textbf{feature vector $\vec{x}^(s)$}} &
\multicolumn{2}{c|}{\textbf{target vector $\vec{y}^(s)$}}\\
\hline
\colrule
$\vec{x}^{(1)}$ & 0 & 0 & 0 & 0 & 0 & 0 &  $\vec{y}^{(1)}$ & 1.15E-05\\
\hline
$\vec{x}^{(2)}$ & 1 & 0 & 0 & 0 & 0 & 0 &  $\vec{y}^{(2)}$ & 1.26E-05\\
\hline
$\vec{x}^{(3)}$ & 0 & 1 & 0 & 0 & 0 & 0 &  $\vec{y}^{(3)}$ & 1.45E-05\\
\hline
$\vec{x}^{(4)}$ & 0 & 0 & 1 & 0 & 0 & 0 &  $\vec{y}^{(4)}$ & 1.53E-05\\
\hline
$\vdots$ & $\vdots$ & $\vdots$ & $\vdots$ & $\vdots$ & $\vdots$ & $\vdots$ &  $\vdots$ & $\vdots$\\
\hline
$\vec{x}^{(62)}$ & 1 & 0 & 1 & 1 & 1 & 1  & $\vec{y}^{(62)}$ & 2.27E-05\\
\hline
$\vec{x}^{(63)}$ & 0 & 1 & 1 & 1 & 1 & 1  & $\vec{y}^{(63)}$ & 2.53E-05\\
\hline
$\vec{x}^{(64)}$ & 1 & 1 & 1 & 1 & 1 & 1  & $\vec{y}^{(64)}$ & 2.71E-05\\
\hline
\end{tabular}
\end{table}

\begin{eqnarray}
E(q) = \sum_{i=1}^6 \sum_{j=1}^6 Q_{ij}q_iq_j = \sum_{i=1}^6 Qiiqi \nonumber \\
 + \sum_{i=1}^5 \sum_{j=i+1}^6 Q_{ij}q_iq_j
\label{eq:two}
\end{eqnarray}
where $q_i$ is a binary variable which takes either 0 or 1, and coefficients $Q_ij$ and $Q_ii$ have real values. Here, the second equality holds because $q_i^2=q_i$. The $Q_ij$ and $Q_ii$  are obtained through the learning process with a Factorization Machine (FM) predictor \cite{rendle_fm_2010} with model equation:
\begin{equation}
\hat{y}(\vec{x}) = \sum_{i=1}^6 w_i q_i + \sum_{i=1}^5 \sum_{j=i+1}^6 (\vec{v}_l\cdot \vec{v}_j)q_iq_j
\label{eq:three}
\end{equation}

\begin{equation}
(\vec{v}_l\cdot \vec{v}_j) = \sum_{f=1}^k v_{if}\cdot v_{jf}
\label{eq:four}
\end{equation}
where $\kappa$ is the dimension of the factorization. Following previous prescriptions \cite{kitai_2020}, $\kappa$ is set to 8 during the learning process to minimize the cost function.

Note that although the unconstrained QUBO model can also be determined by the regression of $Q_{ij}$ and $Q_{ii}$ in equation (1), the advantage of using the FM predictor is that it can reliably fit the parameters (i.e. $w_i$, $\vec{v}_l$ and $\vec{v}_j$ with a small training dataset and prevent the trained model from overfitting as previously shown \cite{kitai_2020}. This prediction model is termed the unconstrained QUBO model. For a thorough discussion on the learning ability of FM, we refer the reader to previous studies \cite{rendle_fm_2010, rendle_2012, rendle_pairwise_2010}.

The performance of the unconstrained QUBO model is evaluated by checking the correlation (i.e., the square of the sample correlation coefficient denoted by $R^2$) between the FC values of all deuterated molecular structures not selected by the unconstrained QUBO model and the quantum chemistry calculations. If $R^2$ is less than a threshold value of 0.95, a randomly chosen $\vec{x}^{(s)}$ possessing one deuterium atom and its reverse feature vector is added to the training data to retrain the unconstrained QUBO model. The procedure repeats by choosing $\vec{x}^{(s)}$ while gradually increasing the number of deuterium atoms (i.e., the bit value of 0) in the feature vector, until $R^2$ surpasses the threshold value.

\subsection{\label{sec:level2}Construction of the Constrained Prediction Model}

After creating an unconstrained QUBO model to accurately predict the FC values, we then build a new constrained QUBO model which includes a penalty term added to the unconstrained QUBO model:
\begin{eqnarray}
QUBO_{constrained} = QUBO_{unconstrained} \nonumber \\
 + \beta_0 QUBO_{penalty}
\label{eq:five}
\end{eqnarray}
where $\beta_0$ is a weighting parameter . The penalty QUBO model is also obtained using the FM predictor with a new dataset corresponding to the feature vector $\vec{x}^{(s)}$. The new dataset has a format similar to that shown in Table 1. The $\vec{x}^{(s)}$ in the new dataset also represent all 64 deuterated molecules and the corresponding target values $\vec{y}^{(s)}$ are generated by
\begin{equation}
y^{(s)} = f(n_i) = (n_i - n_0)^2
\label{eq:six}
\end{equation}
where $n_i$ is the number of deuterium atoms (corresponding to bit value 0) in the i-th of $\vec{x}^{(s)}$ and $n_0$ is a specific integer value with $0 \leq n_0 \leq 6$. Thus, the $\vec{y}^{(s)}$ has a minimum of value 0 when $n_i=n_0$. Note that in order to make the values in $QUBO_{unconstrained}$ and $QUBO_{penalty}$ possess the same order of magnitude, $Q_{ij}$ and $Q_{ii}$ in $QUBO_{unconstrained}$ are scaled by dividing the minimum value of $Q_{ij}$ and $Q_{ii}$ into all $Q_{ij}$ and $Q_{ii}$. Note that in order to guarantee the minimum of $QUBO_{unconstrained}$ has $n_0$ deuterium atoms, the value of $\beta_0$ in Eq (5) is set to 10.

\subsection{\label{sec:level2}Quantum Optimization Methodology}

To perform quantum optimizations on quantum simulators and devices using the VQE and QAOA algorithms, both unconstrained and constrained QUBO models are converted into a spin-based Ising model Hamiltonian (denoted $s_i$) which takes a value of +1 or -1 \cite{ising_1925}:
\begin{equation}
H(s) = \sum_{i=1}^6 h_i s_i + \sum_{j=i+1}^6 J_{ij} s_i s_j
\label{eq:seven}
\end{equation}
where $s_i$ is the Pauli Z operator acting on the \textit{i-th} qubit, and the relation between $q_i$ in eq (1) and $s_i$ is given by Equation 8,

\begin{equation}
q_i = \frac{s_i + 1} {2}
\label{eq:eight}
\end{equation}
which indicates how the classical parameter $q_i$ is replaced by the quantum parameter $s_i$.

In the VQE method, the ground state energy of the $H(s)$ is calculated by minimizing the mean energy
\begin{equation}
E(\theta) = \langle{\psi(\theta) \vert H(s) \vert \psi(\theta)}\rangle
\label{eq:nine}
\end{equation}
where $|\psi(\theta)\rangle$ is a parameterized Ans{\"a}tze
\begin{equation}
\vert \psi(\theta)\rangle =  U(\theta)\vert \psi_0\rangle
\label{eq:ten}
\end{equation}
with $U(\theta)$, the unitary operator of the quantum circuit and $|\psi_0\rangle$, the initial state.

In the QAOA method, the parameterized Ans{\"a}tze state is constructed from $H(s)$ with single qubit Pauli $X$ operators as:
\begin{eqnarray}
\vert \psi_p(\vec{\gamma},\vec{\beta)}\rangle =  e^{-i\beta_p B} e^{-i\gamma_p H(s)} \ldots \nonumber \\
e^{-i\beta_1 B} e^{-i\gamma_1 H(s)} \vert +\rangle^{\bigotimes 6}
\label{eq:eleven}
\end{eqnarray}

\begin{equation}
B = \sum_{i=1}^6 X_i
\label{eq:twelve}
\end{equation}
where $p$ is an integer parameter representing the number of repetitions of the unitary operators.

The $R_y$ heuristic Ans{\"a}tze, which is suitable for calculations on quantum devices, was used in the VQE simulations with a circuit depth equal to 1. The 6-qubit circuit featuring the $R_y$ Ans{\"a}tze has 5 nearest-neighbor CNOT gates and 12 optimization parameters $\theta$ as shown in Figure 3(a). The parameter $\theta$ is updated so that the mean energy (i.e., $E(\theta)$ shown in Eq (9)) decreases in each iteration.

\begin{figure}
\includegraphics{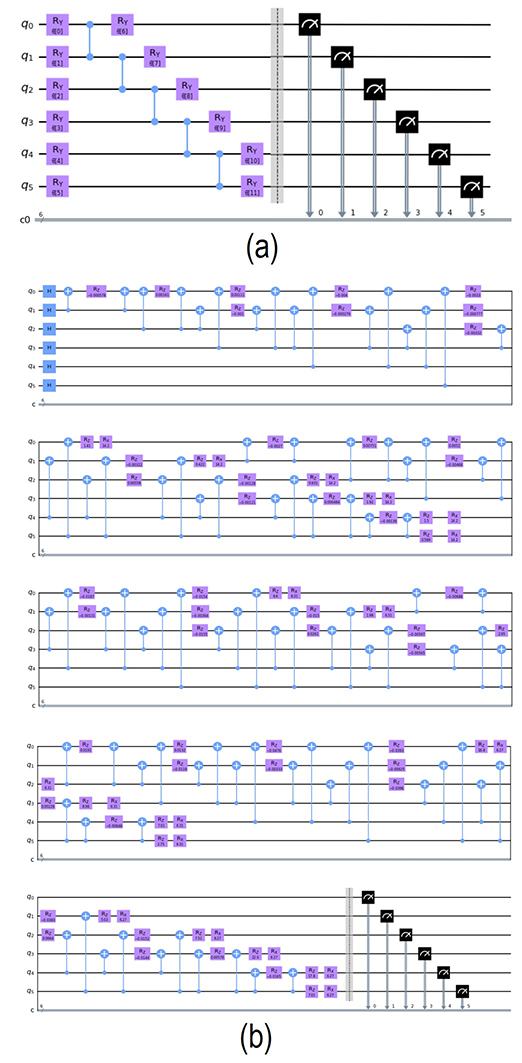}
\caption{\label{fig:fig3}6-qubit circuits for (a) VQE with the $R_y$ Ans{\"a}tze and (b) QAOA with $p=3$.}
\end{figure}

Like VQE, the parameters ($\gamma_i$,$\beta_i (i=1,...,p)$) are updated so that the mean energy (given by $E(\theta) = \langle{\psi_p(\vec{\gamma},\vec{\beta)} \vert H(s) \vert \psi_p(\vec{\gamma},\vec{\beta)}}\rangle$) decreases in each iteration. An example QAOA circuit with $p=3$ is shown in Figure 3(b) which requires 90 CNOT gates and 6 optimization parameters. The advantage of using the QAOA Ans{\"a}tze is that adiabatic dynamics can be used to effectively find the ground state of the Hamiltonian to solve the combinatorial optimization problem \cite{farhi_2014,farhi_2017}.

All quantum optimization calculations were performed with the \texttt{statevector} simulator contained in the Aer module of Qiskit 0.25 \cite{gadi_aleksandrowicz_2019}, as well as on the $ibm\_kawasaki$ 27-qubit quantum device. The \texttt{statevector} simulator uses linear algebra operations to compute the Ans{\"a}tze state and expectation values exactly. The Constrained Optimization BY Linear Approximation (COBYLA) optimizer \cite{powell_1994} was used to update parameters for calculations on both simulator and quantum device.

The readout error mitigation routine in Qiskit was used to mitigate errors occurring when computing ground state energies. In detail, the 64 computational basis states of the 6-qubit system are measured for the purpose of readout error mitigation, then the vectors of probability for the 64 basis states are created for every measurement, and the combination of all 64 vectors resulted in the formation of a $64x64$ matrix from which a calibration matrix was generated by the least-squares fitting. The calibration matrix was applied to the measurements obtained from VQE and QAOA calculations to correct the mean energy value. The parameter values in the VQE algorithm with the $R_y$ Ans{\"a}tze and QAOA are then calculated using the COBYLA optimizer. Six linearly connected qubits were selected to reduce the influence of CNOT errors on the accuracy of results computed on the quantum device. Figure 4 show the architecture of the $ibm\_kawasaki$ device and the quantum circuits comprising the 6 chosen qubits (i.e. [q0,q1,q2,q3,q5,q8]) implementing VQE with the $R_y$ Ans{\"a}tze and QAOA with $p=3$.

\begin{figure*}
\includegraphics{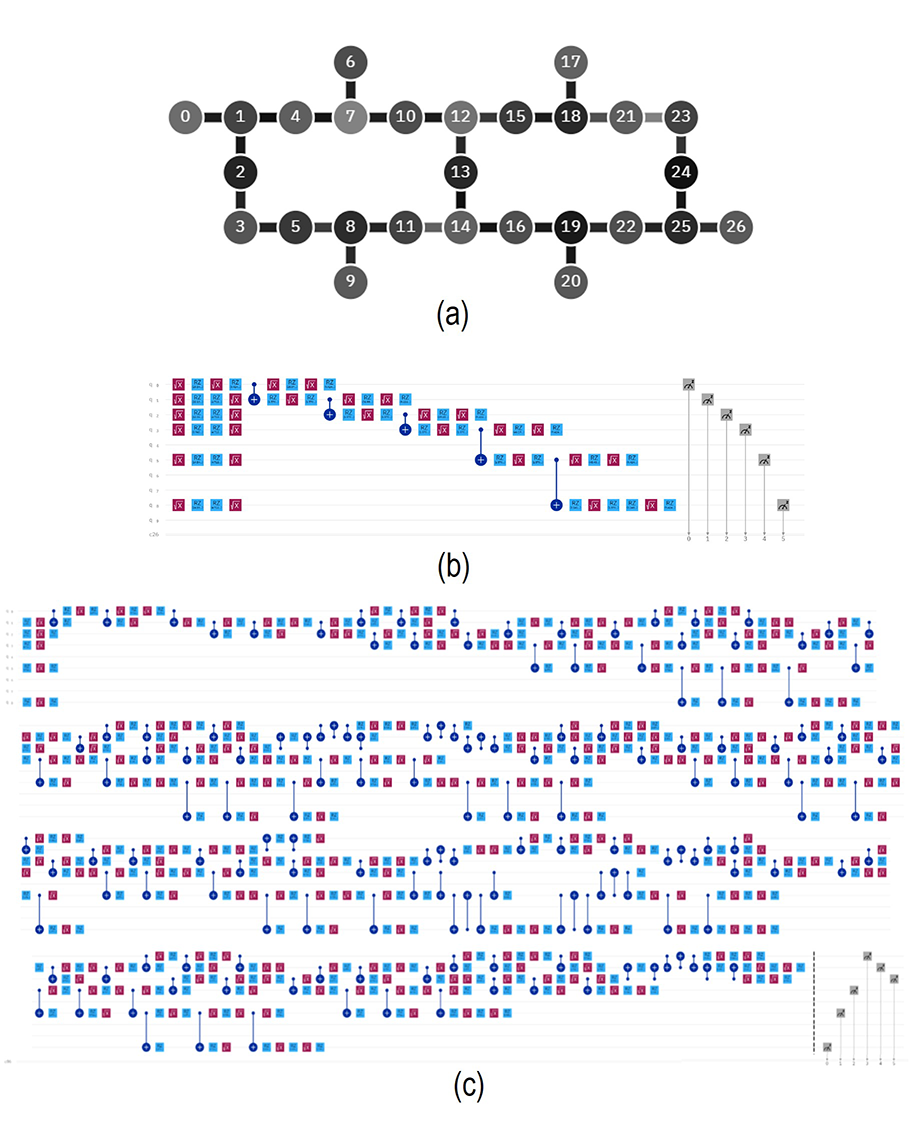}
\caption{\label{fig:fig4}(a) The architecture of the $ibm\_kawasaki$ 27-qubit quantum device (b) quantum circuits implemented on $ibm\_kawasaki$ for (b) VQE and (c) QAOA with $p=3$.}
\end{figure*}

\subsection{\label{sec:level2}Application of the binary search algorithm in optimizations on the quantum device}

Since one of the computational basis states must comprise the optimal bitstring, we can view searching for the optimal bitstring as a problem involving determining the binary values of every qubit. This inspired us to suggest a new approach, illustrated in Figure 5, which applies the binary search algorithm to VQE and QAOA calculations to improve the accuracy on a quantum device. In this approach, VQE and QAOA calculations are first performed on all the 6 qubits with a loose optimization convergence criterion. Using the computed results, the probability of the output of 0 or 1 for every qubit is then calculated as:

\begin{equation}
p_i^\textrm{0 or 1} = \frac {\textrm{counts for 0 or 1 on i-th qubit}} {\textrm{total measurement counts}}
\label{eq:thirteen}
\end{equation}

For qubits with $p_i^\textrm{0 or 1}$ larger than a threshold value denoted by $\delta$, we determine the binary values of the qubits and construct a new Ising model Hamiltonian by omitting the terms of these related qubits. With the new Ising model Hamiltonian in hand, VQE and QAOA calculations are repeated for the remaining qubits with $p_i^\textrm{0 or 1} \leq \delta$ until the binary values of all the 6 qubits are determined. We expect that this approach is more effective for finding the optimal bitstring on a quantum device than the original VQE and QAOA approach due to the following reasons: 1) although noise from the quantum device causes both the accuracy of $p_i^\textrm{0 or 1}$ and the probability of the highest bitstring in the original VQE and QAOA simulations to decrease, the values of $p_i^\textrm{0  or  1}$ are larger than the probability of highest bitstring; 2) leaving out qubits with $p_i^\textrm{0 or 1}>\delta$ causes the number of qubits required for VQE and QAOA calculations to be reduced and consequently the binary values of some qubits are determined with smaller-depth quantum circuits than the original circuits for VQE and QAOA calculations.

\begin{figure}
\includegraphics{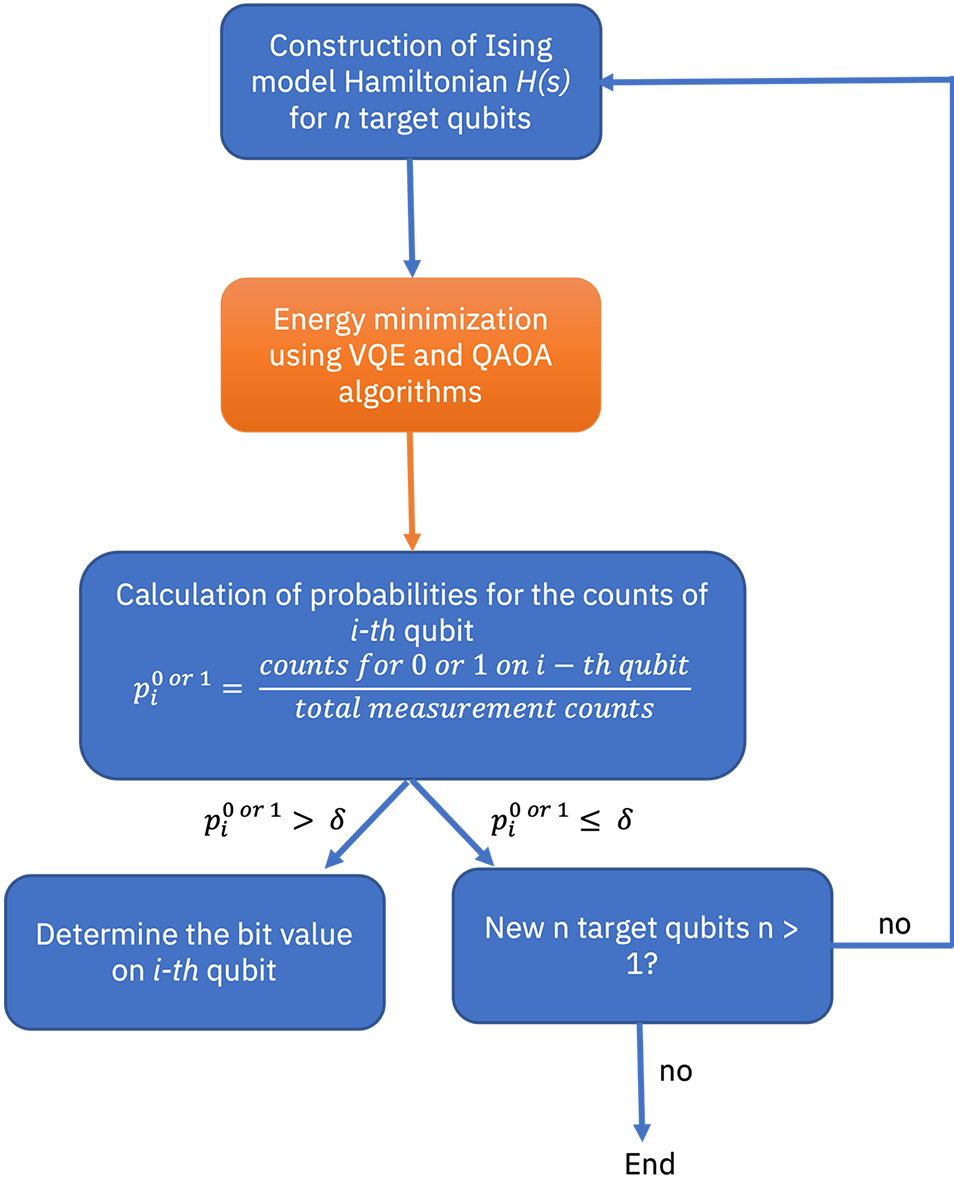}
\caption{\label{fig:fig5} Flowchart showing the binary search algorithm using VQE and QAOA to find the optimal bitstring on quantum devices. Boxes colored blue and orange indicate simulations performed on classical and quantum computers, respectively.}
\end{figure}

\section{\label{sec:level1} Results and Discussion}

\subsection{\label{sec:level2} Quantum Chemistry and Machine Learning}

The ability of the model obtained from the trained FM method to accurately predict FC factors was determined by comparing results predicted by the unconstrained QUBO model with results obtained from ab initio quantum chemistry calculations, as shown in Figure 6. With the smallest training dataset containing 3 deuterated molecules, the unconstrained QUBO model yields FC factors that are 100-1000 larger than results obtained from quantum chemistry results and the correlation between FC factors obtained with the unconstrained QUBO model and the ab initio prediction is 0.032. Increasing the number of deuterated molecules in the training dataset to 11, improves the order of the predicted FC values from the unconstrained QUBO model but the predicted FC values are still approximately an order of magnitude higher than quantum chemistry predictions. Moreover, the use of this larger training dataset results in only slightly better correlation between the QUBO model and ab initio prediction with an  $R^2$ of 0.11. Gratifyingly, the use of 13 deuterated molecules as a training dataset resulted in excellent agreement between the unconstrained QUBO model predictions and the quantum chemistry calculations and improved the $R^2$ to 0.99. These results signify that machine learning can accurately predict the FC factors of $2^n$ deuterated molecules from training data derived from order n quantum chemistry calculations.

\begin{figure*}
\includegraphics{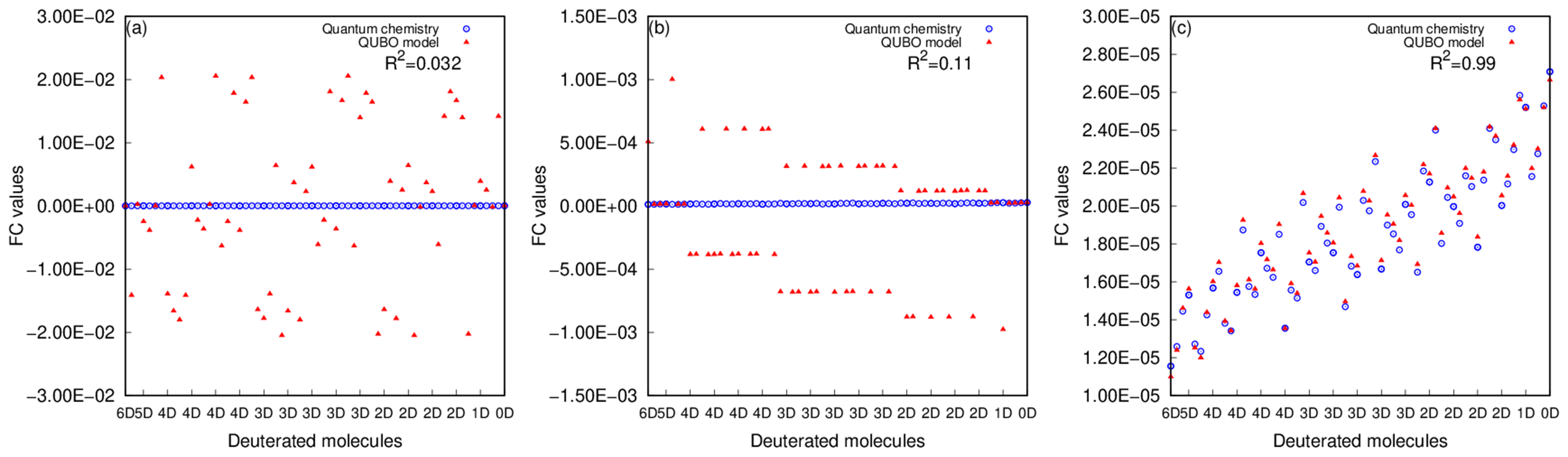}
\caption{\label{fig:fig6} Predicted Franck-Condon (FC) factors for $Alq_3$ molecules labeled with deuterium atoms by \textit{ab initio} calculations and QUBO models trained by the FM predictor with (a) 3, (b) 11 and (c) 13 deuterated molecules. $R^2$ is the square of the sample correlation coefficient between FC factors obtained from \textit{ab initio} calculations and the QUBO models.}
\end{figure*}

These 13 deuterated $Alq_3$ derivatives comprise the non-deuterated molecules with feature vector $\vec{x}^{(s)} = [111111]$ as well as the deuterated molecules containing only one deuterium atom or one hydrogen atom i.e., the one-hot ([000001], [000010], etc.) and one-cold ([111110], [111101], etc.) feature vectors. Because training an unconstrained QUBO model can be viewed as determining the values of entries in a 6x6 matrix, the values of 6 diagonal entries in the matrix are mainly determined by the 6 one-hot vectors in $\vec{x}^{(s)}$. Similarly, the 6 one-cold vectors and [111111] in $\vec{x}^{(s)}$ are used to determine the values of off-diagonal entries. This means that exclusion of some deuterated molecules in the training data will cause the learning process to be more difficult when trying to accurately determine the values in the matrix and emphasizes the need for an accurate unconstrained QUBO model.

Choosing a constrained QUBO model to find an optimally deuterated molecule possessing high desired quantum efficiency requires the constrained QUBO model to be in good agreement with quantum chemistry predictions of FC factors for all deuterated structures. Moreover, the deuterated structure corresponding to the global minimum of the constrained QUBO model must contain the expected number of deuterium atoms.

Figure 7 shows a representative plot of values obtained for an unconstrained QUBO model in comparison with a constrained QUBO model built using equations (5) and (6) with $n_0 = 3$ for the set of deuterated molecules. These results show that the $QUBO_{penalty}$ term has no effect on deuterated molecules possessing 3 deuterium atoms i.e., the values predicted by unconstrained and constrained QUBO models are the same for these molecules and are in good agreement with the FC factors obtained from quantum chemistry calculations. Consequently, the constrained QUBO model can reliably search the global minimum for $Alq_3$ derivatives possessing 3 deuterium atoms. On the other hand, for deuterated $Alq_3$ derivatives with more or less than 3 deuterium atoms, inclusion of the $QUBO_{penalty}$ term in the constrained QUBO model predicts values that are 40\% to 860\% larger than the values of the unconstrained QUBO model. Similar results were also observed for other constrained QUBO models penalized with $n_0=2,4, or 5$. These results indicate that constrained QUBO models can be used to evaluate additional desirable traits, such as stability and low toxicity, by including adequate $QUBO_{penalty}$ terms.

\begin{figure}
\includegraphics{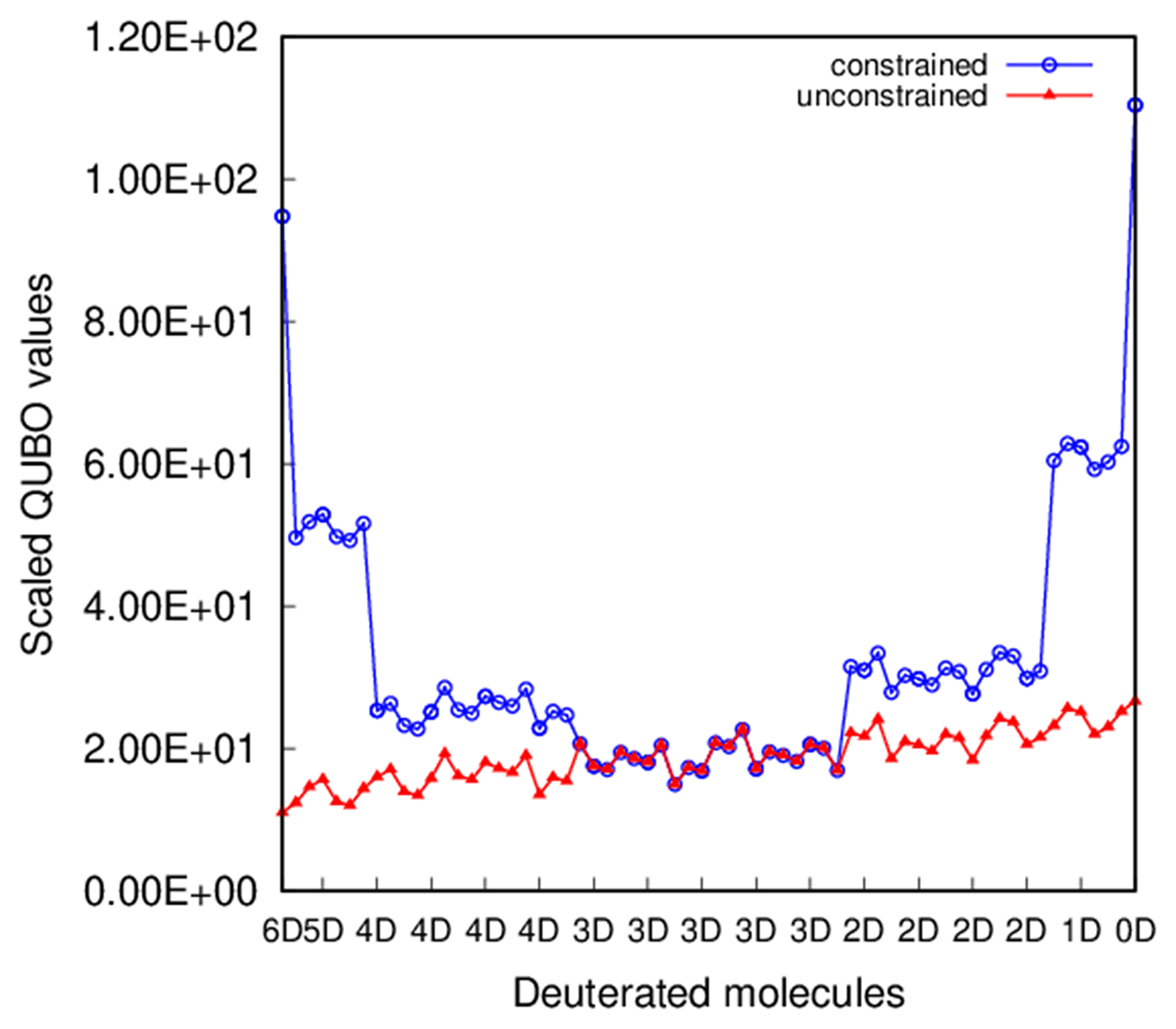}
\caption{\label{fig:fig7}Values of all the deuterated molecules predicted by the scaled unconstrained and constrained QUBO models.}
\end{figure}

\subsection{\label{sec:level2}Simulations with the \texttt{statevector} simulator}

The search for an optimal deuterated molecule can be mapped to a quantum optimization problem whose ground state corresponds to the optimal deuterated molecule by converting the unconstrained and constrained QUBO models described in the preceding section to the unconstrained and constrained Ising Hamiltonian models using equations (7) and (8). The ground state energy and the corresponding bitstrings appearing with highest probability, calculated by the exact eigensolver (i.e., an exact diagonalization of the Ising model Hamiltonian), VQE and QAOA algorithms with the \texttt{statevector} simulator for the unconstrained and constrained Ising models are shown in Table 2.

\begin{table*}
\caption{\label{tab:table2}%
Ground state energies and corresponding bitstrings appearing with highest probability computed using various Ans{\"a}tze with the \texttt{statevector} simulator. Ans{\"a}tze properties comprising the number of CNOT gates and optimization parameters (opt params) are also shown.
}
\begin{tabular}{|p{3cm}|p{1.5cm}|p{4cm}|p{1.5cm}|p{1.5cm}|p{1.5cm}|}
\hline
\multicolumn{1}{|c|}{\textbf{Method}} &
\multicolumn{3}{c|}{\textbf{Ground state}} &
\multicolumn{2}{c|}{\textbf{Ans{\"a}tze}} \\
\hline
\multicolumn{1}{|c|}{\textrm{}} &
\multicolumn{1}{|c|}{\textrm{energy}} &
\multicolumn{1}{|c|}{\textrm{bitstring}} &
\multicolumn{1}{|c|}{\textrm{value}} &
\multicolumn{1}{|c|}{\textrm{CNOTs}} &
\multicolumn{1}{|c|}{\textrm{opt params}}\\
\hline
\multicolumn{6}{|c|}{\textit{Simulations using unconstrained Ising Hamiltonian}}\\
\colrule
exact eigensolver & 10.984 & [000000] (DDDDDD) & 1.00 &  &  \\
\hline
VQE & 10.984 & [000000] (DDDDDD) & 1.00 & 5 & 12 \\
\hline
QAOA $(p=1)$ & 13.170 & [000000] (DDDDDD) & 0.44 & 30 & 2 \\
\hline
QAOA $(p=2)$ & 12.041 & [000000] (DDDDDD) & 0.65 & 60 & 4 \\
\hline
QAOA $(p=3)$ & 11.026 & [000000] (DDDDDD) & 0.97 & 90 & 6 \\
\hline
\multicolumn{6}{|c|}{\textit{Simulations using constrained Ising Hamiltonian}}\\
\hline
exact eigensolver & 14.948 & [100110] (HDDHHD) & 1.00 &  &  \\
\hline
VQE & 14.948 & [100110] (HDDHHD) & 1.00 & 5 & 12 \\
\hline
QAOA $(p=1)$ & 21.870 & [100110] (HDDHHD) & 0.46 & 30 & 2 \\
\hline
QAOA $(p=2)$ & 19.589 & [100110] (HDDHHD) & 0.68 & 60 & 4 \\
\hline
QAOA $(p=3)$ & 19.569 & [100110] (HDDHHD) & 0.78 & 90 & 6 \\
\hline
QAOA $(p=4)$ & 15.388 & [100110] (HDDHHD) & 0.95 & 120 & 8 \\
\hline
\end{tabular}
\end{table*}

VQE with the heuristic $R_y$ Ans{\"a}tze reproduces the ground state calculated with the exact eigensolver using an unconstrained Ising model Hamiltonian. The [000000] bitstring, corresponding to the fully deuterated $Alq_3$ ligand, is obtained with probability of 1.00, indicating that quantum optimization with VQE successfully finds the exact solution for the unconstrained Ising model Hamiltonian. While QAOA converges to bitstrings corresponding to the fully deuterated $Alq_3$ derivative, the probabilities for this selection only range from 0.44 to 0.97 increasing with the values of $p$ and not quite matching the probability of 1.00 observed with VQE. Similarly, the converged energy obtained from QAOA calculations decreases with increasing values of $p$. These improved behaviors are presumably caused by the increased accuracy of the description of the time-evolution operator with the increasing integer value of the parameters of $p$ \cite{farhi_2014}.

Results produced with the constrained Ising model Hamiltonian are largely like those found with the unconstrained Ising model Hamiltonian: VQE produces the same accuracy and QAOA results also improve with increasing values of $p$. However, it was observed that to have a result comparable with the exact solution, $p=4$ was needed to converge the ground state with a probability of 0.95 with QAOA. Such behavior may be due to the energy differences between the unconstrained and constrained Ising model Hamiltonians shown in Figure 8. While 4 excited states are obtained for the unconstrained Ising model Hamiltonian, 19 excited states are obtained for the constrained Ising model Hamiltonian, and the gaps between the ground state and the first excited state for the unconstrained and constrained Ising model Hamiltonians are 1.00 and 0.40, respectively. Because more excited states and a smaller energy band gap are obtained for the constrained Ising model Hamiltonian, the optimization is more sensitive to the influence of the excited states, and it is more difficult to obtain the ground state for the QAOA calculation using the constrained Ising model Hamiltonian with $p=3$ than a similar calculation using the unconstrained Ising model Hamiltonian. Nevertheless, since the most probable bitstring from all the QAOA calculation is the same as the optimal bitstring obtained via the exact eigensolver, these results can be useful when searching for find the optimal deuterated molecule.

\begin{figure}
\includegraphics{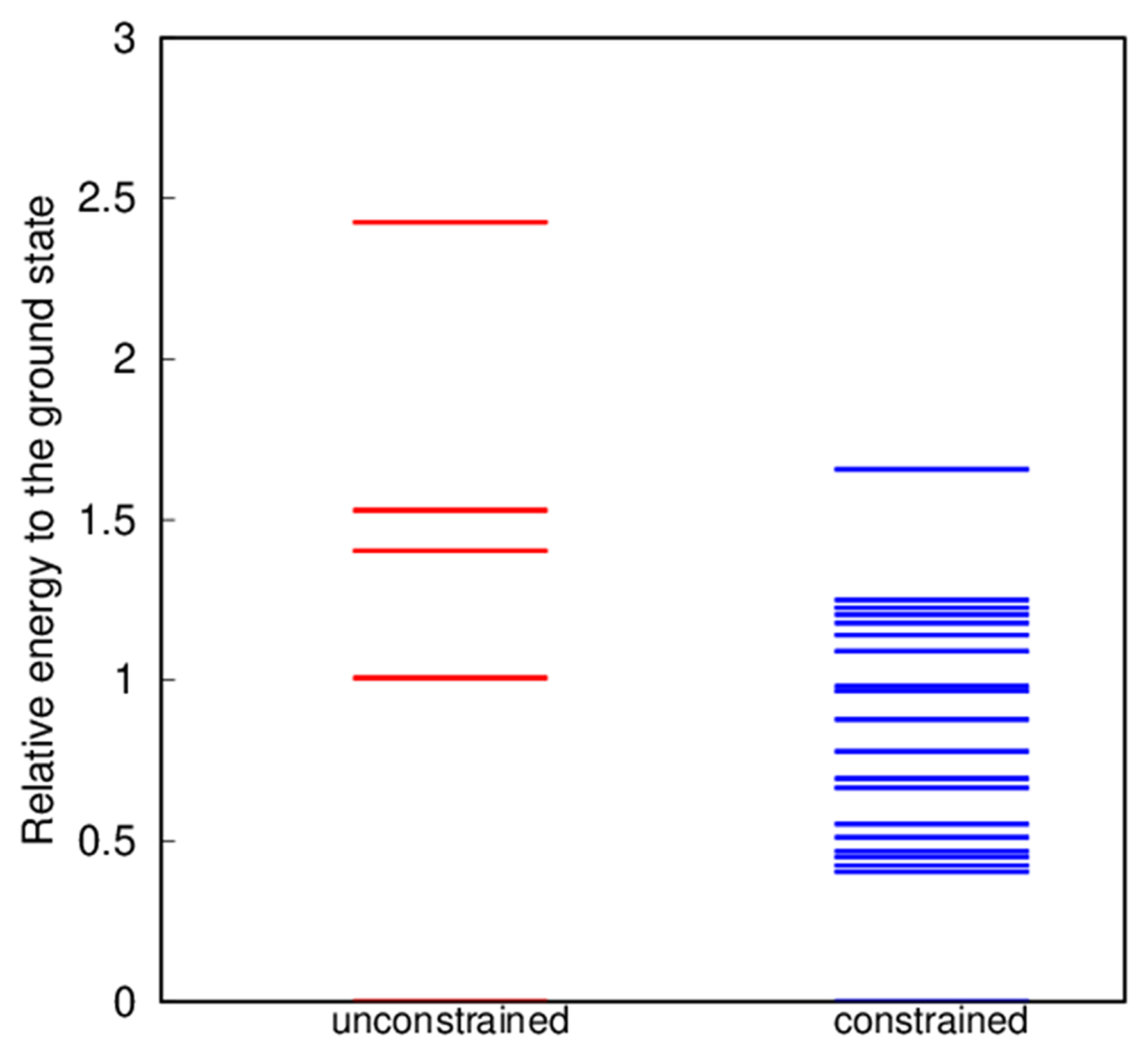}
\caption{\label{fig:fig8} Relative energies of the eigenstates with respect to the ground state in the unconstrained and constrained Ising model Hamiltonians within a range of 2.5 mH.}
\end{figure}

Table 2 also shows the number of CNOT gates in the quantum circuits and the number of optimization parameters used in the VQE and QAOA algorithms. This data was used to determine whether the optimization algorithms could be implemented on the $ibm\_kawasaki$ device for simulations involving 6 qubits. The quantum circuits for VQE calculations involving the use of the $R_y$ Ans{\"a}tze possess 5 CNOT gates, whereas 30 CNOT gates are required for QAOA simulations with $p=1$ and this linearly increases with the value of $p$. Note that since the $R_y$ heuristic Ansätze employs nearest-neighbor coupling between the qubits in the so-called light-cone structure, by using the linearly connected 6-qubit subsection of $ibm\_kawasaki$, the number of CNOT gates would not change for when the circuit is implemented. In contrast, since QAOA does not have such a convenient coupling structure, an overhead of 3x~4x more qubits would be needed for implementing the circuit on the quantum device. Since the CNOT error rate can negatively affect the accuracy of the computed results, simulation with the VQE Ans{\"a}tze on $ibm\_kawasaki$ is expected to provide a higher accuracy than provided by QAOA.

VQE requires 12 optimization parameters, while only 2 parameters are needed for QAOA with $p=1$; this value linearly increases with the value of $p$, and thus even with $p=4$ the number of optimization parameters is 8 which is still smaller than the number of parameters required for VQE. QAOA is expected to search these parameters more capably on a classical computer than VQE since a smaller number of optimization parameters infers better trainability and, thereby, a reduction in the number of iterations required for convergence of the energy.

\begin{figure*}
\includegraphics{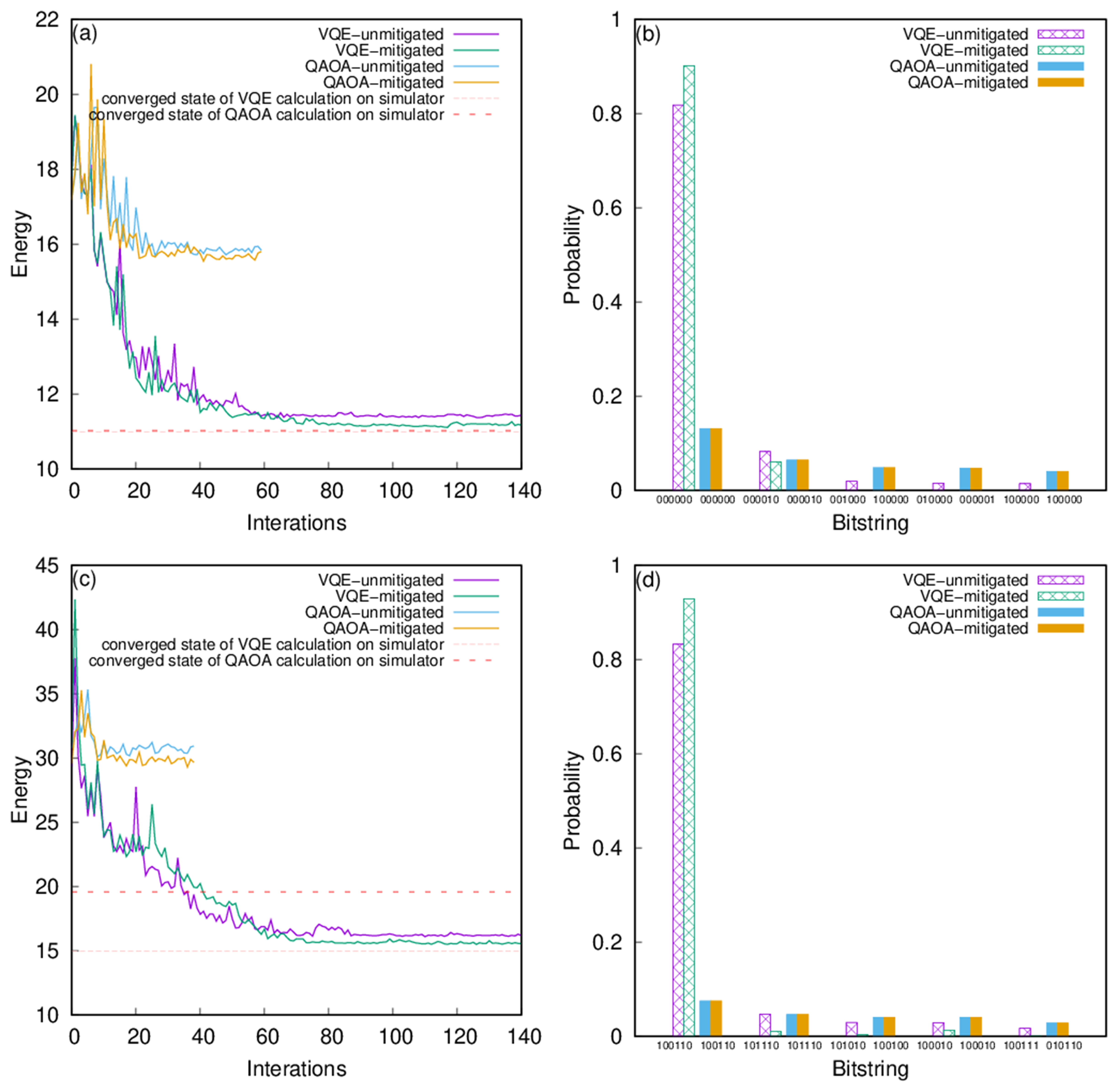}
\caption{\label{fig:fig9} The iteration steps in VQE and QAOA (b) $ibm_kawasaki$ with the (a) unconstrained and (c) constrained Ising model Hamiltonians, and the top five high probability bitstrings of the converged state in the VQE and QAOA calculations the (b) unconstrained and (d) constrained Ising model Hamiltonians.}
\end{figure*}

The energy minimization and the success probability of the converged state in the VQE and QAOA calculations using the constrained Ising Hamiltonian with and without the readout error mitigations are shown in Figure 9(c-d). The deviation of ground state energy and the highest bitstring probability in VQE calculations on the quantum device from predictions obtained with the quantum simulator are very similar to VQE calculations using the unconstrained Ising Hamiltonian as shown in Figure 9(a-b). On the other hand, the deviations observed for QAOA calculations using the constrained Ising Hamiltonian are much larger than those with the unconstrained Ising Hamiltonian. For example, the energy deviations from the QAOA calculations using the constrained and unconstrained Ising Hamiltonian with readout error mitigation are 10.18 and 4.64, respectively, and their corresponding highest probabilities are 0.08 and 0.14. The lower accuracy of QAOA with the constrained Ising Hamiltonian may be due to the energy differences of the Hamiltonians as shown in Figure 8 and discussed in the preceding section. Overall, that fact that VQE can predict the optimal bitstring with probability larger than 0.9 on the quantum device is quite satisfactory. Although the success probability obtained from QAOA calculation is one order lower than those from VQE, the predicted results can still be useful for finding the optimally deuterated $Alq_3$ derivative.

The five highest probability bitstrings were also investigated for VQE and QAOA calculations and are shown in Figures 10(b,d). Bitstrings with probabilities ranging from the 2nd highest to 5th highest occurrences mostly differ from the dominant bitstring by only a single qubit. For example, as shown in Figure 9(b) for VQE calculation with the unconstrained Ising Hamiltonian the highest probability bitstring comprises the [000000] configuration and the next four bitstrings comprise the [000010], [001000], [010000], [100000] configurations. Consequently, the probability of the correct value 0 on qubits q1, q2, q3, q4 and q6 is above 0.96. The value is much higher than the probability of the dominant bitstring (i.e. [000000]) which is about 0.90. Moreover, we also found that the probability of the correct output on every qubit is quite different. For example, among the top five highest probability bitstrings, the probability of the correct value $0$ on qubit \textit{q5} is about 0.90 which is much lower than the probability on qubits \textit{q1, q2, q3, q4} and \textit{q6}. This is a consequence of the fact that the impact of noise on each of these qubits is variable.

\subsection{\label{sec:level2}Simulations on $ibm\_kawasaki$}

Simulations using VQE and QAOA with $p=3$ using the unconstrained Ising model Hamiltonian were performed on $ibm\_kawasaki$. The iteration steps (i.e., the number of updates of the parameters) versus the energy values, as well as the five greatest probability bitstrings of the converged state, are shown in Figure 9(a-b). To mitigate the measurement errors from the quantum device, we have applied readout error mitigation to ground states computed with VQE and QAOA. VQE without readout error mitigation yields energies that are about 0.43 higher than the results from the exact values calculated on the simulator. The bitstring appearing with the highest probability is the same as that obtained on the simulator but the probability is lowered by 0.18. The difference of the energy and the success probability are improved to about 0.21 and 0.10, which highlights the importance of the inclusion of readout error mitigation in VQE calculations on the quantum device. Although the most probable bitstring obtained via QAOA is the same as that predicted by VQE, the deviations of the success probability and the converged state from those obtained with QAOA on the simulator are approximately an order of magnitude larger than those obtained with VQE. This is caused by the fact that QAOA involves many more gate operations than VQE as shown in Figure 4. For example, there are 5 and 156 CNOT gates, which significantly limits the accuracy of quantum algorithms, in the VQE and QAOA quantum circuits, respectively. We also found that the application of readout error mitigation in QAOA results only in marginal improvement of the ground state energy and the value of highest probability of bitstring. These results suggest that errors such as decoherence noise is the main cause for the deviation of QAOA results from the ideal.

The preceding results inspired us to use the binary search algorithm in the VQE and QAOA calculations to reduce the effect of noise from the quantum device by systematically searching the correct output binary value on every qubit as shown in Figure 5, instead of directly searching for the optimal bitstring. The results of VQE and QAOA calculations using the binary search for the unconstrained Ising Hamiltonian are shown in Figure 10(a-b).

\begin{figure*}
\includegraphics{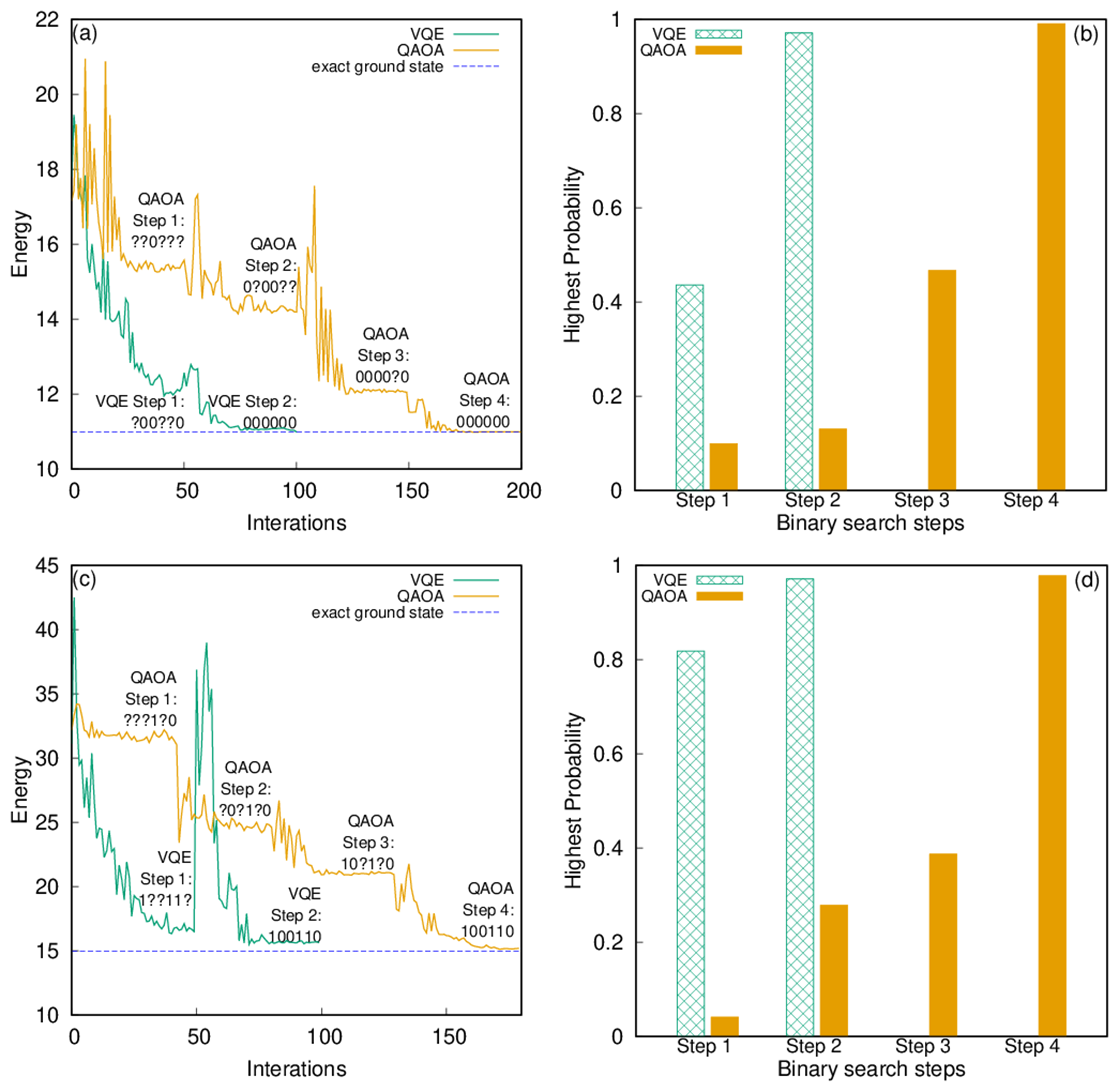}
\caption{\label{fig:fig10} The iteration steps in VQE and QAOA (b) $ibm\_kawasaki$ with the (a) unconstrained and (c) constrained Ising model Hamiltonians, and the top five high probability bitstrings of the converged state in the VQE and QAOA calculations the (b) unconstrained and (d) constrained Ising model Hamiltonians.}
\end{figure*}

Applying the binary search algorithm to determine binary values of target qubits improves the accuracy of ground state energies compared to both the original VQE and QAOA calculations. The original VQE and QAOA calculations using the unconstrained Ising Hamiltonian yields energies that are about 0.21 and 4.64 higher than the exact values; this gap can be improved to about 0.08 and 0.01 by using the binary search algorithm. Notably, the improvement of the ground state prediction observed for QAOA is more significant than observed for VQE. This is due to the fact that the decreased number of target qubits causes the number of gate operations to decrease by a factor of 10-100 in the QAOA calculations whereas with VQE this decrease is only by a factor of 2-3. However, it must be noted that many more iterations are required to converge QAOA combined with binary search in comparison to VQE. Because the QAOA calculation for 6 qubits is much more sensitive to the effect of noise from the quantum device than the VQE calculation, the first application of binary search in the QAOA calculation can only leave 1 qubit out of 6 qubits whereas 3 qubits can be leaved out in the VQE calculations. Consequently, the VQE calculation can converge to the ground state using two binary searches, whereas 4 binary searches are needed for the QAOA calculation, as shown in Figure 10(a). The improvement of the values of the highest probability for the VQE and QAOA calculations with the binary search is also observed. Without the binary search, the values of the highest probability are 0.90 and 0.14 for the original VQE and QAOA calculations, respectively. Calculations using the binary search improves these values to 0.97 and 0.99 respectively, signifying the effectiveness of using the binary search to find the optimal bitstring on quantum device. Moreover, we note that the value of the highest probability from the QAOA calculations using the binary search on the quantum device is somewhat higher than the values obtained on the \texttt{statevector} simulator, as shown in Figure 10(b) and Table 2. This is caused by the fact that the number of terms in the Ising Hamiltonian decreases when the binary search algorithm is applied to QAOA simulations. Consequently, the energy eigenstates have smaller variance, and the converged state becomes closer to the exact ground state. Similar results are also found in the VQE and QAOA results using the constrained Ising Hamiltonian as shown in Figure 10(c-d).

\section{\label{sec:level1}Conclusions}

A combined quantum chemistry, machine learning and quantum optimization approach has been developed for the molecular design of high-efficiency, synthesizable deuterated $Alq_3$ emitters. Quantum chemistry calculations of FC factors provide a consistent picture of the emission quantum efficiency of deuterated $Alq_3$ emitters. Based in part upon these quantum chemistry studies, an unconstrained Ising Hamiltonian was constructed with a factorization machine (FM) predictor and its ability to reproduce the property of emission quantum efficiency has been validated. Moreover, a synthetic constraint was introduced based on the fact that the synthesizability of the $Alq_3$ emitters is inversely proportional to the number of deuterium atoms in deuterated $Alq_3$ emitters. A constrained Ising Hamiltonian was constructed by adding the additional synthetic constraint to the unconstrained Ising Hamiltonia. The ground state of the constrained Ising Hamiltonian results in a deuterated $Alq_3$ molecule which meets the desiderata of both emission quantum efficiency and synthesizability.

VQE and QAOA calculations with the constrained and unconstrained Ising Hamiltonians are performed on the \texttt{statevector} simulator and IBM quantum devices. Calculations performed with VQE and QAOA algorithms on the \texttt{statevector} simulator accurately converge to the ground state and successfully find the optimal deuterated $Alq_3$ molecule with probability larger than 0.95. These results show that the VQE and QAOA algorithms can be reliably used to generate lead deuterated $Alq_3$ molecules possessing desirable emission quantum efficiency and are synthesizable.

Results from VQE and QAOA calculations on IBM quantum devices show that noise significantly limits the accuracy of energy values of converged state. As a consequence, the bitstring probability corresponding to the optimal deuterated $Alq_3$ molecule decreases to 0.83 and 0.075 for VQE and QAOA calculations, respectively. The inclusion of readout error mitigation in the VQE calculations increases the probability by 0.07, while only a marginal improvement of the probability was observed in QAOA calculations. To obtain more reliable quantum optimization results from simulations on the quantum device, a new scheme which utilizes the binary search algorithm in the VQE and QAOA calculations was developed to systematically search for the correct output binary value on every qubit. Both VQE and QAOA calculations provide the optimal bitstring with a probability of 0.97 when this scheme is used.

Beyond $Alq_3$, the application of the binary search algorithm results in reliable VQE and QAOA calculation accuracy on quantum device, highlighting opportunities for the use of quantum device in the discovery of deuterated OLED emitters such as novel iridium and platinum complexes. More broadly, leveraging a combined quantum chemistry, machine learning and quantum optimization approaches into a discovery workflow paves the way for another dimension in the generation and optimization of lead molecules in the arena of material informatics.

\section{\label{sec:level1}Acknowledgments}

Q.G., M.S., T.K., H.Y., H.K., S.T. and N.Y. acknowledge support from MEXT Quantum Leap Flagship Program Grant Number JPMXS0118067285.

\bibliographystyle{apsrev4-2}
\bibliography{deuterated_oled}

\end{document}